# Substitution effect of Zn and Cu in $MgB_2$ on $T_c$ and structure

S.M.Kazakov, M.Angst, J.Karpinski*

Solid State Physics Laboratory, ETH, 8093 Zürich, Switzerland

I.M. Fita, R.Puzniak

Institute of Physics, Polish Academy of Sciences, Warsaw, Poland

*Corresponding author, ph: 0041 1 633 22 54, karpinski@solid.phys.ethz.ch

**Abstract**

The investigation of Zn substitution in $MgB_2$ polycrystalline samples shows that about 0.1 Zn can be substituted in the structure of Mg deficient samples while it can not be substituted in stoichiometric $MgB_2$. As a result of the isovalent Zn substitution into the Mg position of Mg deficient samples the *a* and *c* lattice constants increase by about 0.17% and 0.2% respectively. Susceptibility measurements show a slight lowering of $T_c$ by about 0.5 K for 0.05 Zn, which decreases to 0.2 K for 0.1 Zn substitution. Investigation of the pressure derivative of $T_c$ shows $dp/dT_c$ = -0.15 K/kbar independent of Zn content. Substitution of Cu leads to multiphase samples without any changes of lattice constants, which indicates that Cu does not enter the structure of $MgB_2$. The Cu substitution broadens the superconducting transition considerably, while the onset remains unchanged.

**Introduction**

The recent discovery of superconductivity at 39K in $MgB_2$ by Akimitsu et al. [1] has stimulated world wide excitement. During the first month after the discovery dozens of papers have been published concerning the synthesis of polycrystalline samples and the investigation of basic physical properties. The transition temperature is close to the limit suggested theoretically for phonon mediated superconductivity. Therefore, a question arises regarding the mechanism of superconductivity in this system. The boron isotope effect measured by Bud'ko et al. [2] indicates that $MgB_2$ is a phonon-mediated superconductor. According to several authors the high transition temperature is probably due to the low mass of boron, which yields high phonon frequencies. This is supported by band calculations by Kortus et al. [3] which show a strong electron-phonon interaction. A different theoretical model, the "hole superconductivity", where superconductivity originates from the undressing of hole carriers in almost full bands, was proposed by Hirsch [4]. He predicted a hole carrier conductivity, opposite to other metal diborides with the same crystal structure, which show a negative Hall coefficient. An and Pickett [5] suggest conventional phonon-mediated superconductivity driven by $\sigma$ band holes. In fact measurements of the Hall coefficient on polycrystalline $MgB_2$ samples performed by Kang et al. [6] show a positive Hall coefficient, which indicates hole carriers. Tunneling spectroscopy experiments performed by several groups on polycrystalline samples (Sharoni et al. [7], Rubio-Bollinger et al. [8], Karapetrov et al. [9]) show a good fit to the BCS model with phonon mediated superconductivity. The model by Hirsch [4] predicts a strong increase of $T_c$ upon shrinking the lattice constant *a*, which should lead to a positive pressure effect on $T_c$. However, high pressure studies by Lorenz et al. [10] show a negative pressure coefficient of $T_c$.

In order to test various models substitutions leading to hole and electron doping are crucial. Several such experiments concerning substitutions on the Mg site have been published. Both hole doping by a substitution of $Li^{1+}$ (Zhao et al. [11]) and electron doping by a substitution of $Al^{3+}$ (Slusky et al. [12]) for $Mg^{2+}$ led to a decrease of $T_c$.

It is very interesting to investigate variations of the structure as a function of doping and pressure. From the existing experiments we can see that high hydrostatic pressure has an anisotropic influence on the *a* and *c* lattice constants. According to Jorgensen et al. [13] the compression along the *c* axis is 64% larger than along the *a* axis, which is in line with the weaker Mg-B bond. As a result of the compression $T_c$ decreases. Thermal compression along the *c* axis is about twice the one along the *a* axis.

Substitution of Al leads, apart from doping with electrons, to the compression of the structure. It may be caused by a smaller size of Al than Mg ($r_i$(Al)=0.5 Å, $r_i$(Mg)=0.65Å, $r_M$(Al)=1.43 Å, $r_M$(Mg)=1.6 Å, where $r_i$ and $r_M$ – are the ionic and metallic radii, respectively). The compression is anisotropic and for an Al content of x=0.1 leads to a structural instability. The rate of compression along the *c* axis is about twice as much as along the *a* axis.

What is the result of squeezing the structure in the B-B plane? Shortening of the B-B distance should increase the coupling in the plane. The substitution of the smaller Li ion (0.6 Å) for Mg(0.65 Å) leads to the decrease of the lattice constant *a* without changing the lattice constant *c* [11]. Here we have two effects overlapping each other: doping with holes and an increase of coupling due to a shortening of the B-B distance. As a net result $T_c$ decreases.

It will be interesting to study the influence of an isovalent substitution leading only to a modification of the structure without changing the total charge carrier concentration. We have tried two kinds of such substitutions, namely Zn and Cu. According to Sun [14] Cu substituted $MgB_2$ samples have a $T_c$ onset at 49K. The x-ray investigation of their sample shows as a main phase $Cu_2Mg$. According to the authors the *a* and *c* lattice constants decreased, although both atomic and ionic radii of Cu are larger than those of Mg.

## Experimental

$Mg_{1-x}Zn_xB_2$ and $Mg_{1-x}Cu_xB_2$ samples with x=0, 0.05, 0.1, 0.2 were prepared by solid state reaction in flowing argon. Two series of precursors for the synthesis of $Mg_{1-x}Zn_xB_2$ have been prepared:

(1) precursor with a nominal composition $Mg_{1-x}Zn_xB_2$ for x = 0, 0.05, 0.1, 0.2.

(2) precursor with excess of Mg of the composition $Mg_{1.1-x}Zn_xB_2$ for x = 0, 0.05, 0.1.

As starting materials we used Mg powder (99.8%), Zn powder (99.9%), Cu powder (99,9%) and amorphous boron (99,99% pure). Powders were mixed and pressed into pellets. Then pellets were placed in a vanadium container closed with a cup and fired in a tube furnace under pure Ar gas. The samples were heated for one hour at 600°C, one hour at 800°C, and one hour at 900°C. No traces of Mg or Zn appeared outside the crucible. The phase analysis of the resulting samples was performed using a STOE Stadi P powder X-ray diffractometer with a Cu $K_{\alpha 1}$ radiation. The data were collected over 2θ range from 30° to 70° with a 0.02° step.

## Results

As a result of series (1) x-ray pure $Mg_{1-x}Zn_xB_2$ samples have been obtained for x up to 0.1. Small amounts of $MgB_4$ and MgO were found as impurities in the sample with x=0.2. Figure 1 shows the position of (002) and (110) Bragg reflections for different Zn-content (x=0, 0.05, and 0.1). The position of both peaks shifts towards smaller angles with Zn substitution. This result indicates that the lattice parameters increase as the nominal Zn content increases. The lattice constants for $Mg_{1-x}Zn_xB_2$ calculated by a least-square fitting to the position of 8 reflections are presented in Fig.2 (as squares). Substitution of 0.1 Zn results in the increase of the *a* axis by 0.17% and of the *c* axis by 0.2%. Structural investigations show that for x=0.2 the lattice constants do not change in comparison with those for x=0.1, which suggests that the solubility limit has been reached already for x=0.1. We may therefore assume that the composition of the x=0.1 sample is $Mg_{0.9}Zn_{0.1}B_2$. The lattice constants of the x=0 sample are slightly smaller than those published in the literature [13,14], which indicates that the Mg content is lower than stoichiometric due to the partial evaporation of Mg at high temperature. The deficiency of Mg makes a partial substitution of Zn possible.

As a result of series (2) $MgB_2$ samples have been obtained. X-ray investigations show that lattice constants does not change with the variation of the Zn content in the precursor. Figure 2 shows *a* and *c* lattice constants (as triangles) which are constant in the error limits. The lattice constants correspond to stoichiometric $MgB_2$ published recently [12,13]. These results indicate that Zn does not enter the structure and that the substitution of Zn is possible only in nonstoichiometric samples with Mg deficiency. If enough Mg is available, stoichiometric $MgB_2$ without substitution is formed.

As a result of Cu substitution multiphase samples have been obtained. The X-ray powder pattern shows a mixture of $MgB_2$ and other phases, namely $MgCu_2$, MgO and $MgB_4$. With increasing Cu content in the starting mixture, the amount of impurities increases considerably. The lattice constants of Cu-doped $MgB_2$ are the same as for non-substituted samples, which indicates that Cu does not enter the structure of $MgB_2$.

Magnetic measurements have been performed with a SQUID magnetometer. The temperature dependence of the magnetization in an external field of H=1Oe has been recorded for both zero field cooled and field cooled conditions (see Fig. 3,4).

First samples in both series for x=0 composition are stoichiometric $MgB_2$ and Mg deficient $Mg_{1-x}B_2$ where x≈0.1. Surprisingly the $Mg_{1-x}B_2$ sample has $T_c$=38.6K, while $MgB_2$ has slightly lower $T_c$=38.3K. This suggests that the stoichiometric sample is not optimally doped.

The interesting result is that samples with stoichiometric composition $MgB_2$ and $Mg_{0.9}Zn_{0.1}B_2$ have the same effective critical temperature $T_c$=38.3K, the same total charge and the same lattice constant *a*, while the lattice constant *c* differs between them by about $\Delta c$ = 0.005 Å. This can be an indication that increasing the *c* lattice parameter without changing total charge does not influence $T_c$. Recently Yildirim et al. [15] reported that there seems to be a close correlation between the anharmonicity of the $E_{2g}$ in-plane phonon modes, i.e., vibration of boron ions in opposite directions along the *x* (or *y*) axis, with the Mg ions stationary, and the B-B bond length ($d_{BB}$). For $MgB_2$ $d_{BB}$ is 1.764 Å, significantly stretched from its optimal value of 1.65 Å in elemental planar boron, probably due to repulsive interactions between Mg and B ions. This explains the unusual anharmonicity and observed high $T_c$ in $MgB_2$ [15]. The above may indicate that the increase of $T_c$ in $MgB_2$ may be obtained rather by expansion of $d_{BB}$ than by changes in the *c* axis lattice constant.

On the other hand, as can be seen in Fig.3, the sample with a 0.05 Zn substitution has $T_c$ about 0.5 K lower than unsubstituted sample. Surprisingly, a substitution of 0.1 Zn does not decrease $T_c$ more, on the opposite, $T_c$ increases by about 0.3 K and thus finally is 0.2 K lower than for a non-substituted sample. A substitution of 0.2 Zn does not change $T_c$ more, but the width of the transition becomes slightly larger. This is because the solubility limit is reached and additional phases may appear. The non-monotonic variation of $T_c$ was reproduced in a second measurement. We have observed already a similar effect by an isovalent substitution of Sr for Ba in $Y(Ba_{1-x}Sr_x)_2Cu_4O_8$ [16].

The possible explanation can be that three different effects overlap. They are atom size disorder, caused by the introduction into the lattice of atoms with different atomic radius and the expansion of the *a* and c axis. It is known that atom size disorder can decrease $T_c$ [17]. As one can see in Fig.2, in the first step of 0.05 Zn substitution the *c* axis expands by about 0.16. The expansion of the *a* axis is much lower, namely 0.05%. In the second step of 0.1 Zn substitution the expansion of the *c* axis is smaller, only 0.07%, while the expansion of the *a* axis is 0.12%. As we have mentioned before, variations of the *c* axis does not seem to influence $T_c$. If the expansion of the *a* axis has a positive effect on $T_c$, and the disorder introduced by the substitution has a negative effect on $T_c$ we can expect the results observed. If the above interpretation of our results is correct, it is difficult to reconcile it with the model of hole superconductivity, as that model predicts a decrease of $T_c$ upon increasing the *a* axis [4], although indirect effects, such as a doping by inter-band charge transfer, cannot be ruled out.

Magnetic measurements of $Mg_{1-x}Cu_xB_2$ samples show (Fig.4) the same $T_c$ onset as for unsubstituted samples, but a considerably broader transition due to the multiphase sample, which supports x-ray results indicating that Cu does not enter the structure.

In order to proove what is the influence of expansion/contraction of the *a* and *c* lattice parameters on $T_c$ high pressure experiments can be helpful. The eventual influence of the *c*-axis should be easy visible.

**High pressure experiment**

The effect of hydrostatic pressure of up to 11kbar on $T_c$ of Zn substituted samples was determined from $M(T)$ measurements performed on PAR Model 4500 vibrating sample magnetometer. The pressure was built up in the sample chamber before putting it into the cryostat and the pressure affecting the sample during the measurements at low temperatures was checked by measuring the pressure shift of $T_c$ of a calibrating Sn sample.

As a result we found that the variation of $T_c$ with pressure is linear, the same for Zn substituted and unsubstituted samples and corresponds to $dp/dT_c$ = -0.15(1)K/kbar. Pressure has a negative effect on $T_c$ and the value of $dp/dT_c$ is the same as the one observed previously in unsubstituted samples [10].

According to Jorgenson et al. [13] at high pressure the contraction of the *c*-axis is 64% larger than the one of the *a*-axis, what suggests that the contraction of *c* axis can play an important role.

On the other hand comparison of $MgB_2$ and $Mg_{0.9}Zn_{0.1}B_2$ samples, which have the same $T_c$, the *a*-lattice parameter and charge doping level, but a different *c* – parameter suggests that the variation of the *a*-parameter is more important than the *c*-parameter.

In order to decide which parameters can positively influence $T_c$, investigations of uniaxial contraction pressure derivative $(dp/dT_c)_{x,z}$ are necessary. This will be possible from measurements of thermal expansion on single crystals.

**Conclusions**

As a result of 0.1 isovalent Zn substitution into Mg position of Mg deficient samples the *a* and *c* lattice constants increase by about 0.17% and 0.2%, respectively. Susceptibility measurements show a slight lowering of $T_c$ by about 0.5 K for 0.05 Zn, which decreases to 0.2 K for 0.1 Zn substitution. The possible explanation of the above results can be the overlap of the effects of atom size disorder caused by the incorporation of atoms with different atomic radius into lattice (depressing $T_c$) and the expansion of the *a* axis (or B-B bond, increasing $T_c$) and *c* axis (no big impact on $T_c$). Our results indicate that an increase of $T_c$ in $MgB_2$ may be obtained rather by expansion the distance between boron atoms, $d_{BB}$, than by changes in the *c* axis lattice constant.

**Acknowledgement**

This work was supported by the Swiss National Science Foundation and by the Polish State Committee for Scientific Research (KBN) under Contract No. 2 P03B 095 14.

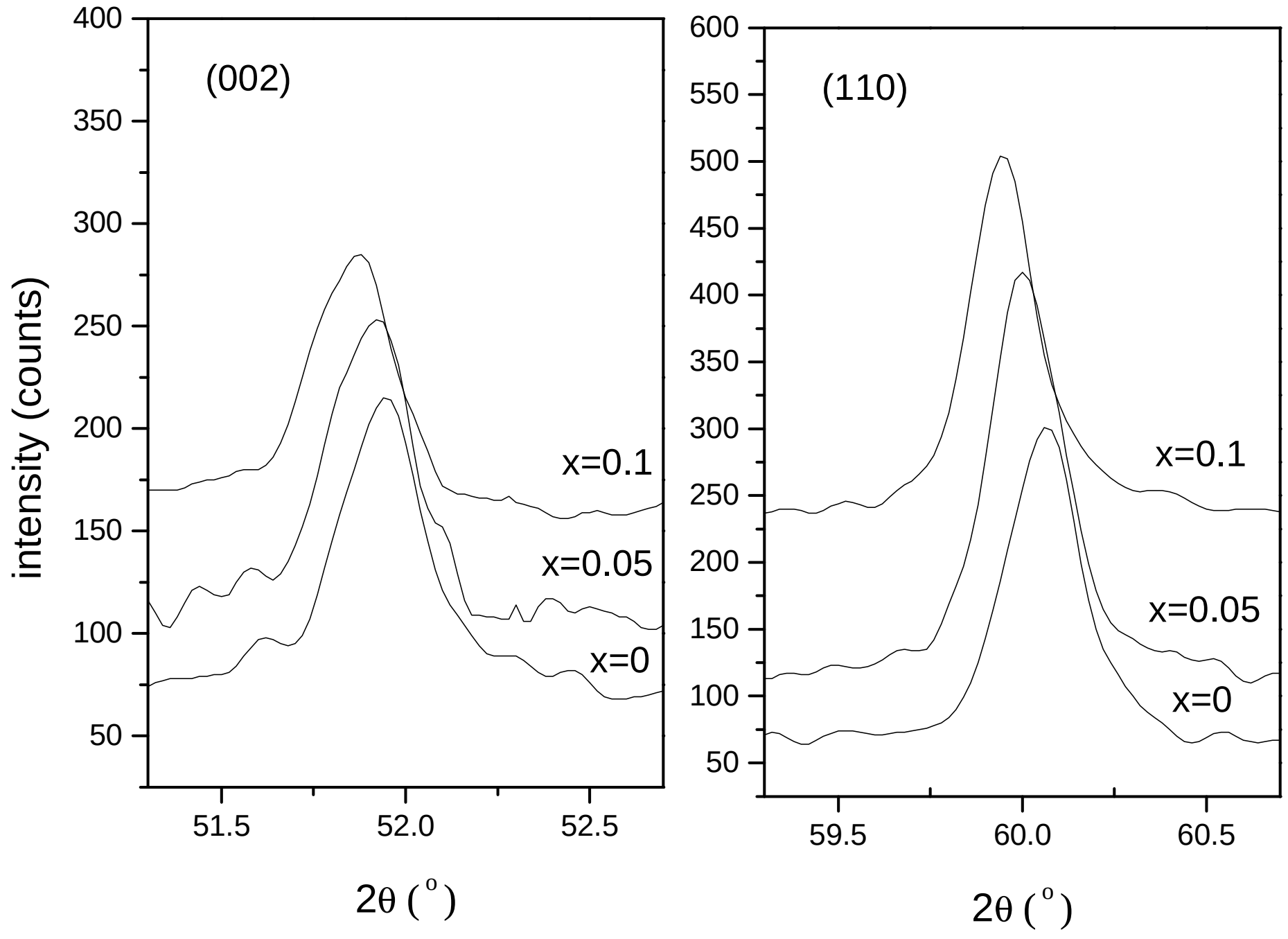

Fig.1. The position of (002) and (110) Bragg reflections for different Zn-content (x=0, 0.05, and 0.1).

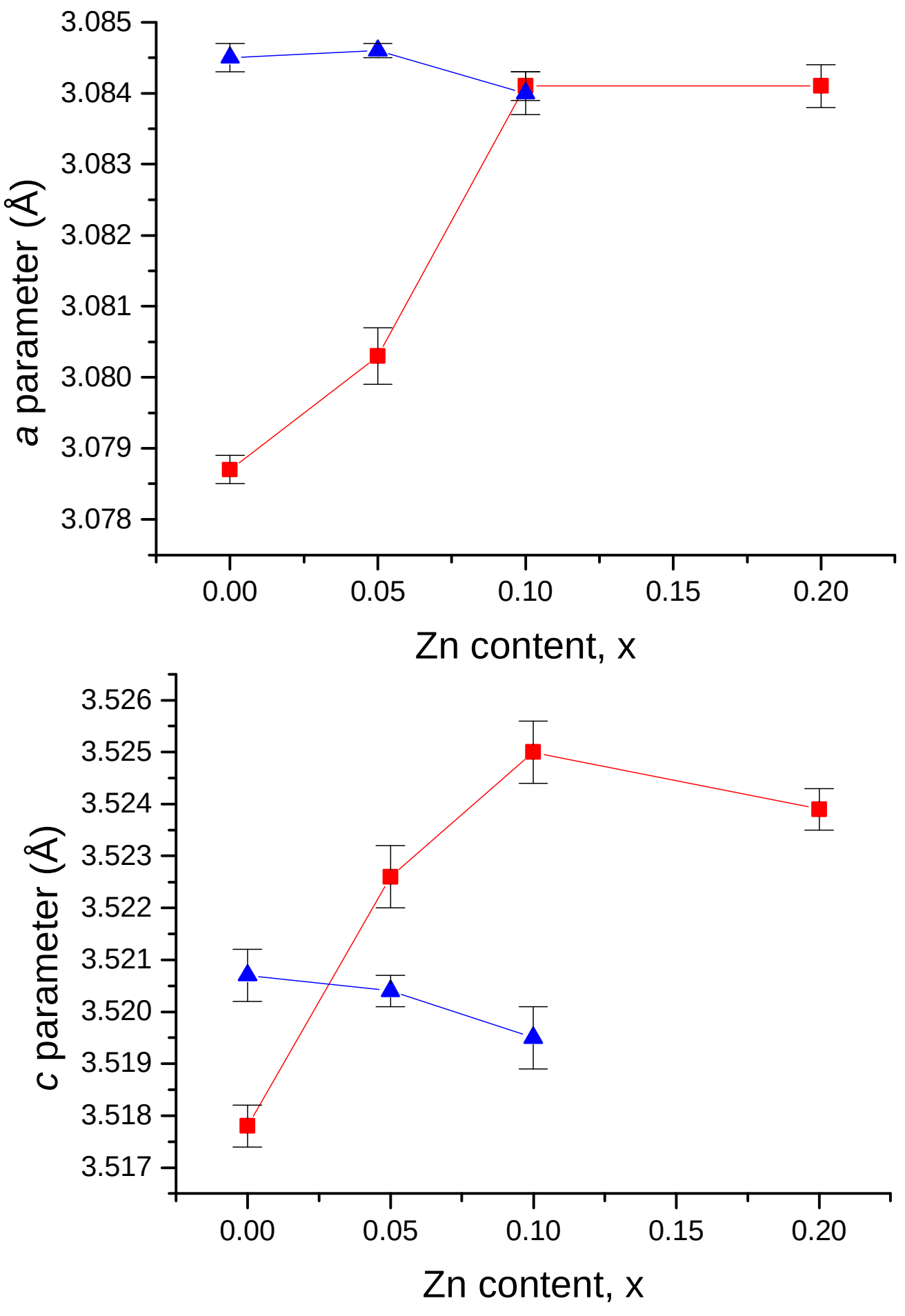

Fig.2. Lattice constants of $Mg_{1-x}Zn_xB_2$. Samples obtained using precursor of composition $Mg_{1-x}Zn_xB_2$: ■, samples obtained using precursor of composition $Mg_{1.1-x}Zn_xB_2$ : ▲. For the x value above 0.1 additional phases appear in the sample.

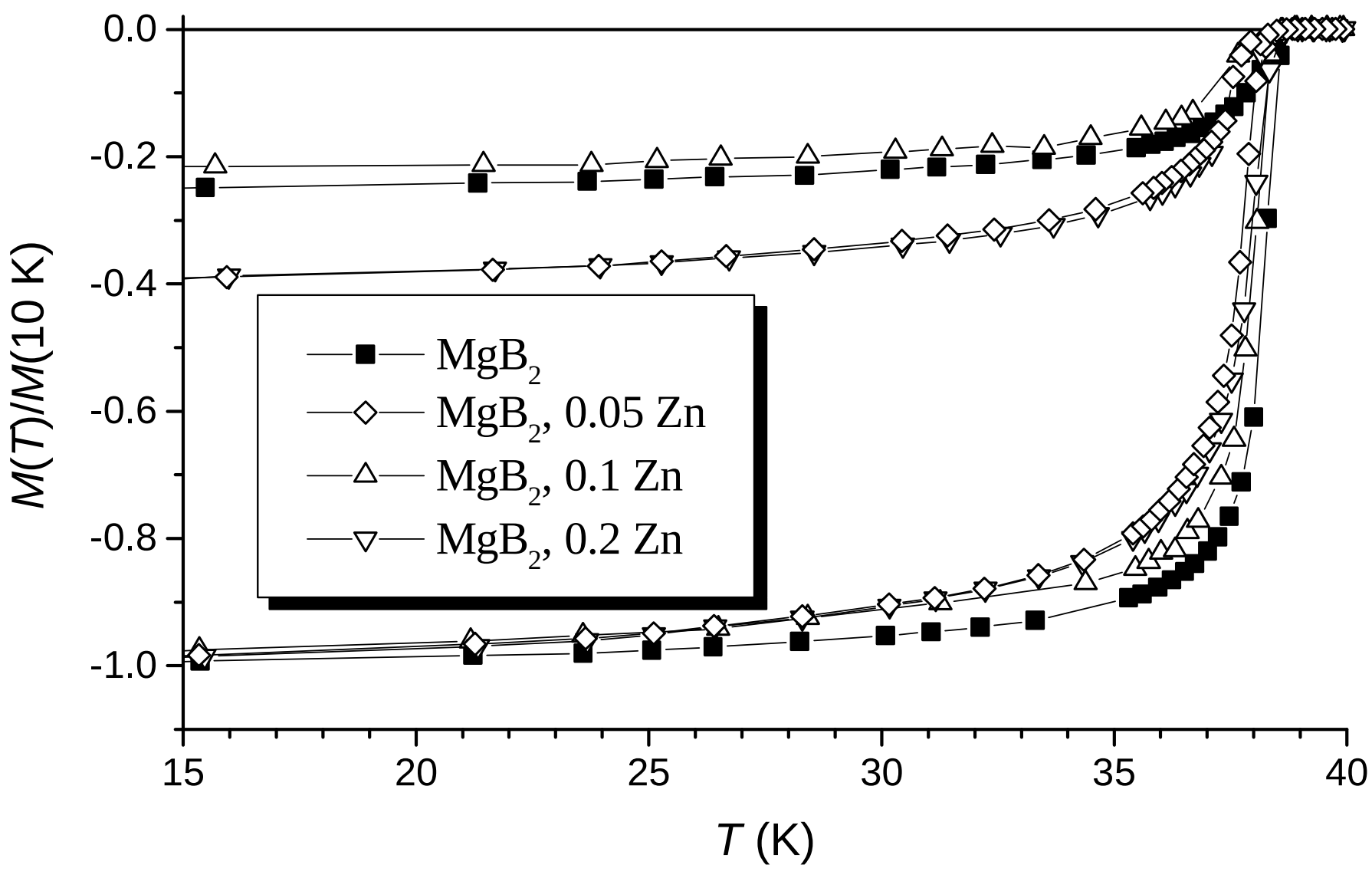

a)

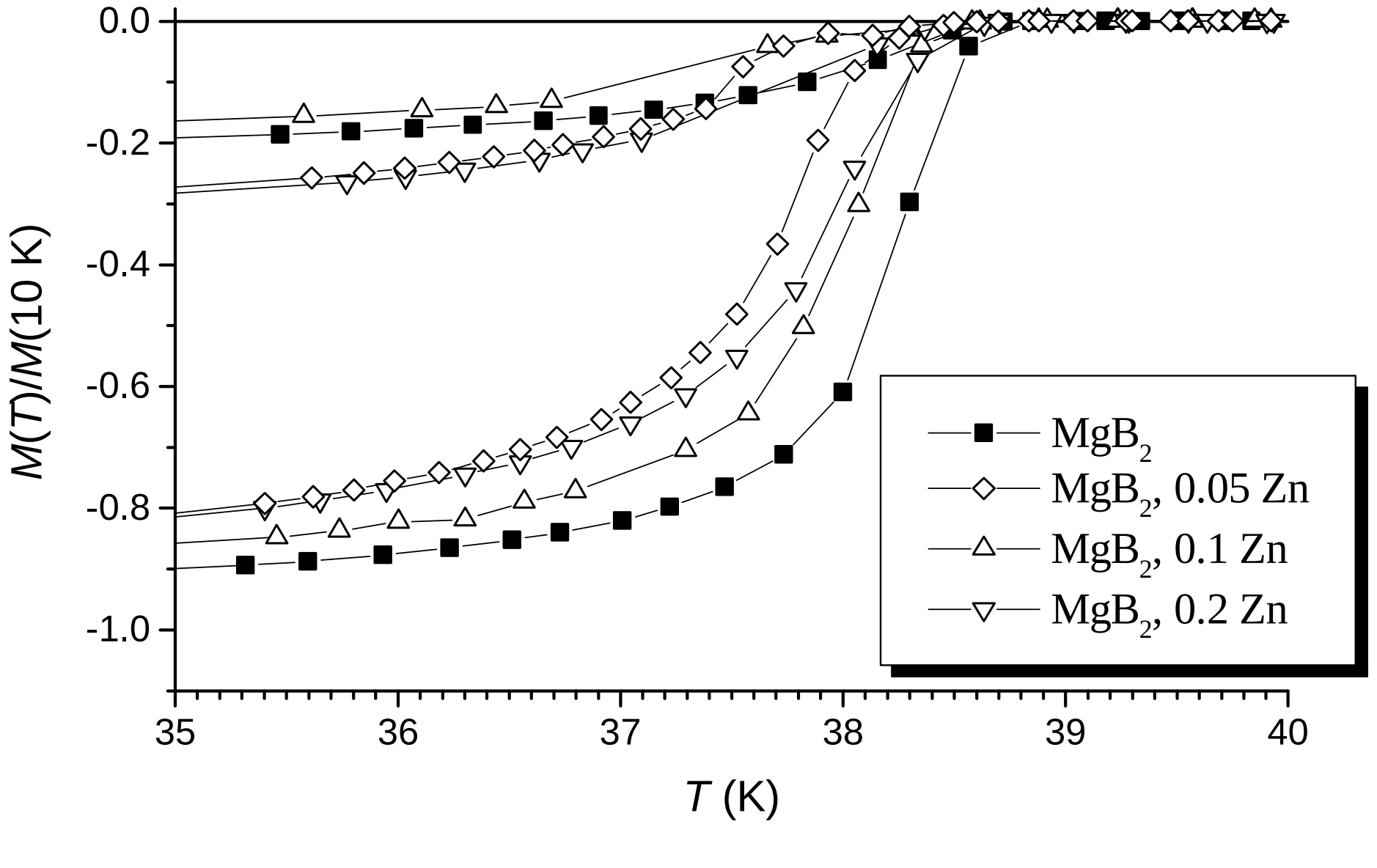

b)

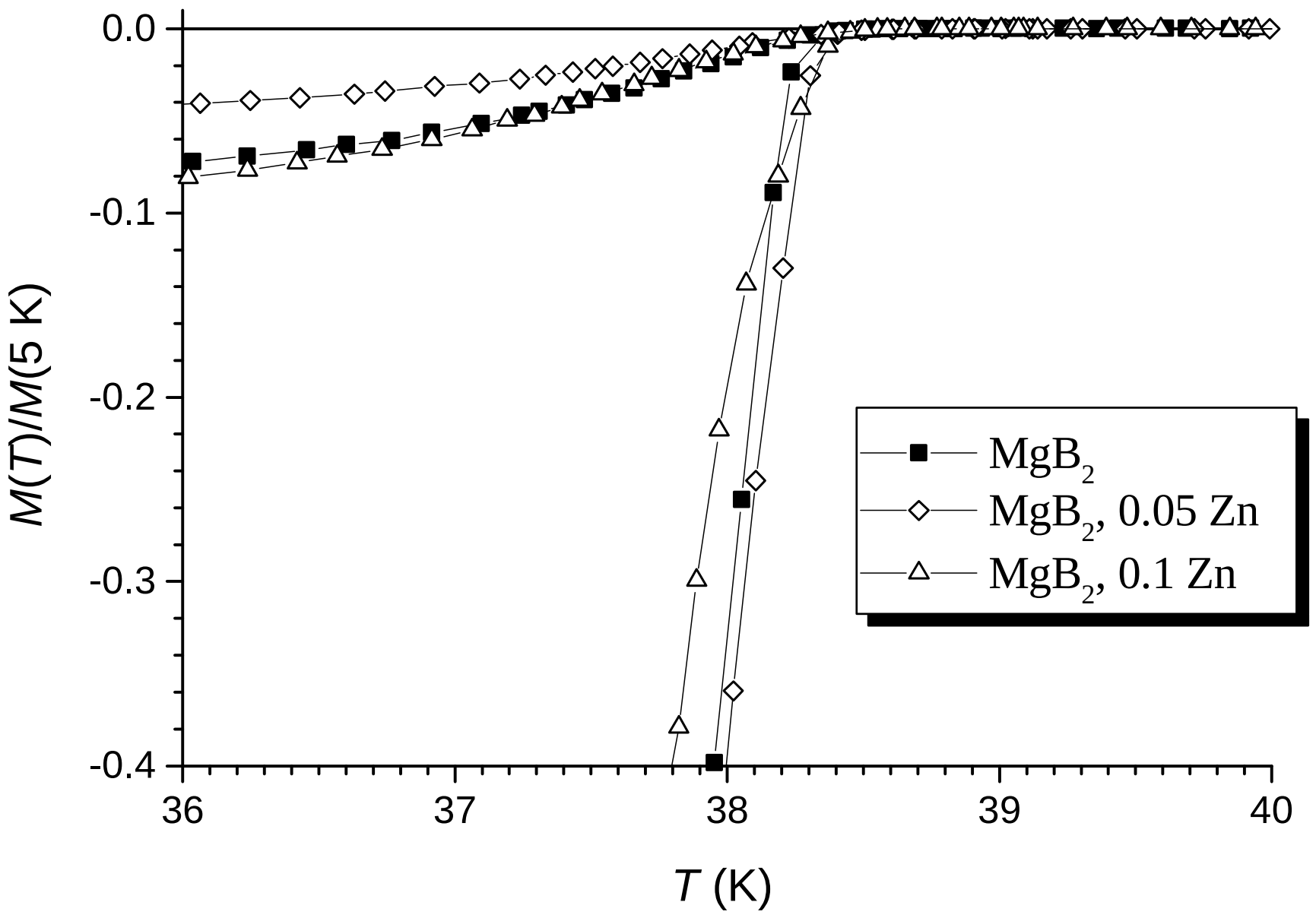

c)

Fig.3. Susceptibility measurements of $Mg_{1-x}Zn_xB_2$ samples: a) whole curve, b) the region of the transition temperature. Samples obtained from the precursor $Mg_{1.1-x}Zn_xB_2$: c).

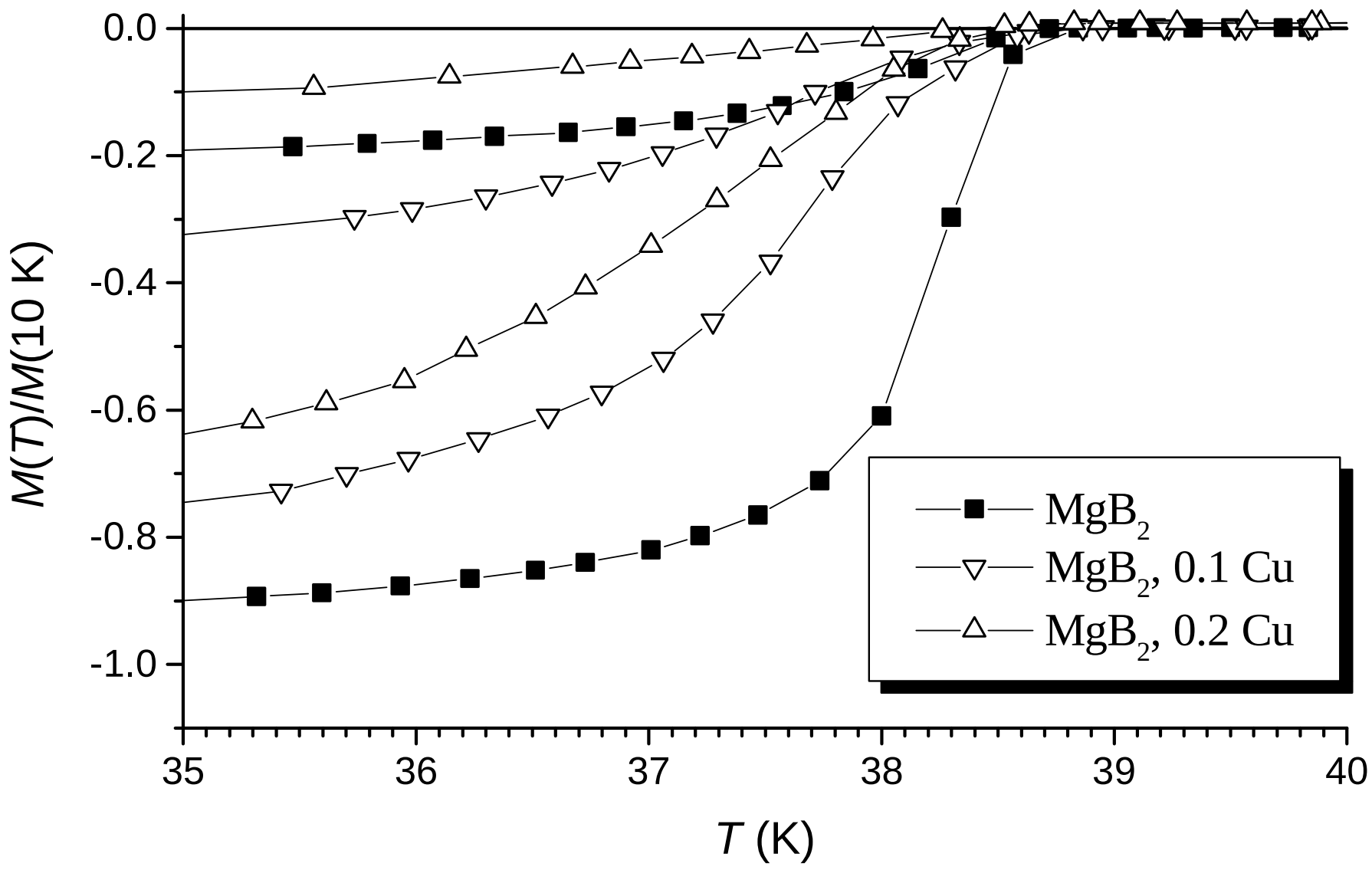

Fig.4. Susceptibility measurements of $Mg_{1-x}Cu_xB_2$ samples.

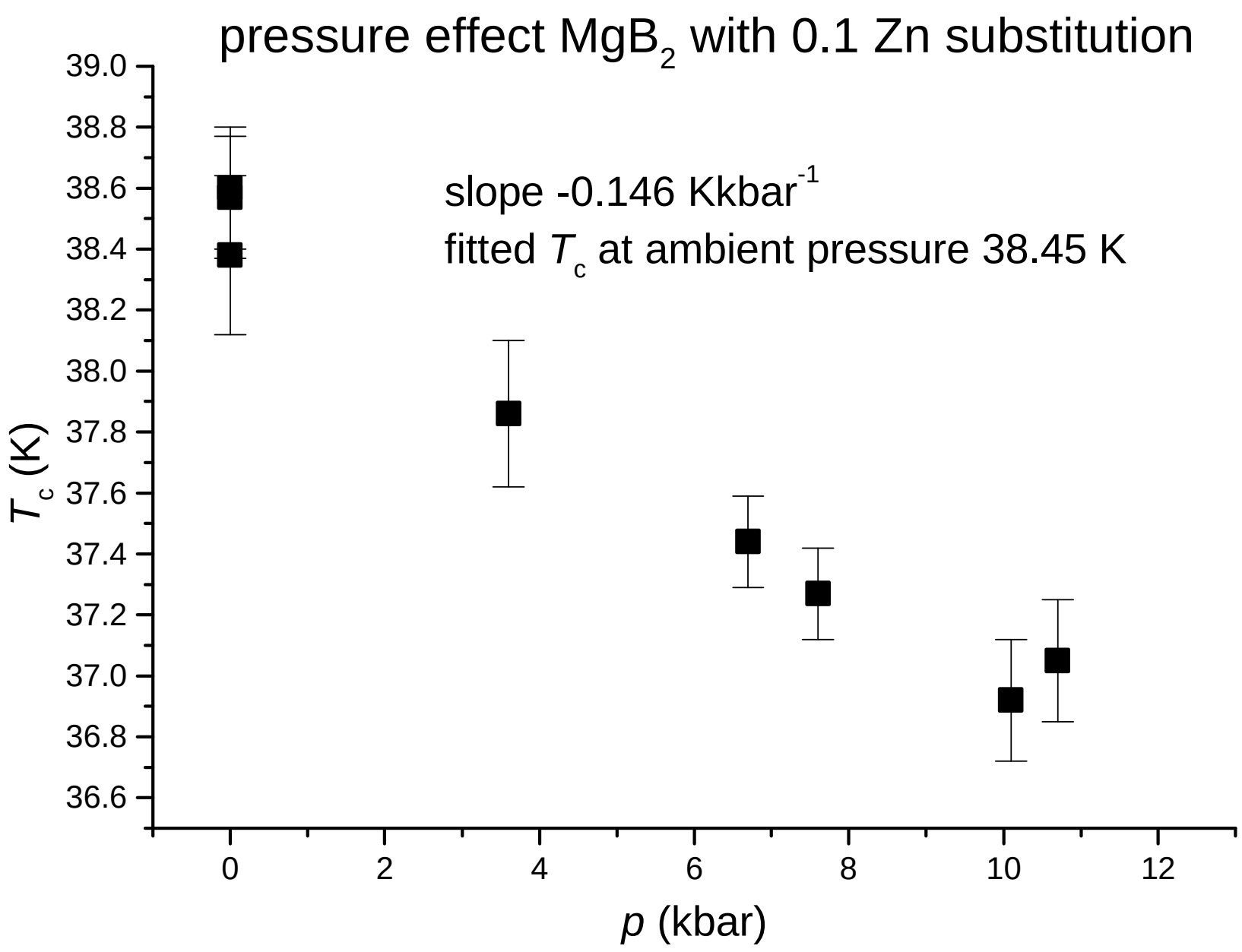

Fig.5. Influence of hydrostatic pressure on $T_c$.